\documentclass[10pt,letterpaper]{article}
\usepackage{opex3}
\usepackage{cite}

\begin{document}

\title{Modelocking and femtosecond pulse generation in chip-based\\ frequency combs}

\author{Kasturi Saha$^1$, Yoshitomo Okawachi$^1$, Bonggu Shim$^1$, Jacob S. Levy$^2$, Reza Salem$^3$, Adrea R. Johnson$^1$, Mark A. Foster$^1$, \\Michael R. E. Lamont,$^{1,2,4}$ Michal Lipson$^{2,4}$, \\and Alexander L. Gaeta$^{1,4}$}

\address{$^1$School of Applied and Engineering Physics, Cornell University, Ithaca, NY 14853 \\ $^2$School of Electrical and Computer Engineering, Cornell University, Ithaca, NY 14853 \\ $^3$PicoLuz, LLC, 10335 Guilford Road, Jessup, MD 20794 \\ $^4$Kavli Institute at Cornell for Nanoscale Science, Cornell University, Ithaca, NY 14853}

\email{a.gaeta@cornell.edu} 



\begin{abstract}
We investigate simultaneously the temporal and optical and radio-frequency spectral properties of parametric frequency combs generated in silicon-nitride microresonators and observe that the system undergoes a transition to a mode-locked state. We demonstrate the generation of sub-200-fs pulses at a repetition rate of 99 GHz. Our calculations show that pulse generation in this system is consistent with soliton modelocking. Ultimately, such parametric devices offer the potential of producing ultrashort laser pulses from the visible to mid-infrared regime at repetition rates from GHz to THz. 
\end{abstract}

\ocis{(190.4380) Nonlinear optics, four-wave mixing; (190.4390) Nonlinear optics, integrated optics; (320.7120) Ultrafast phenomena.} 


\section{Introduction}

Development of ultrashort pulse sources has had an immense impact on condensed-matter physics, biomedical imaging, high-field physics, frequency metrology, telecommunications, nonlinear optics, and molecular spectroscopy \cite{Shelton,Ehlers,Ye,Diels}. Numerous advancements of such sources have been made\cite{Cundiff,Schroder,Martinez,Haus,Xiao,Benedict}, and in recent years there has been development of compact solid-state \cite{Bartels,Pekarek} and semiconductor-based systems that enable high-repetition-rate pulse generation \cite{Lorenser,Klopp,Hoffman,Rodriguez,Akbulut}. However, it remains a challenge to create a highly compact, robust platform capable of producing femtosecond pulses over a wide range of wavelengths, durations, and repetition rates. 

A novel approach to developing such pulse generating platform is suggested by recent research efforts on the generation and characterization of microresonator-based optical frequency combs \cite{KippenbergReview} that are generated via ultra-broadband parametric oscillation based on four-wave mixing (FWM) pumped by a continuous-wave (cw) field. In this process, two pump photons are initially converted to a signal and idler photon pair and the FWM gain is enhanced by the cavity geometry and transverse confinement, which leads to cascaded parametric oscillation, enabling the generation of a broadband comb. In addition, silicon-nitride (Si$_3$N$_4$) microresonators have emerged as a highly promising platform for generating parametric combs since the cavity dispersion and the free spectral range (FSR) (i.e., the comb spacing) can be independently tuned. We have previously shown that parametric oscillation \cite{Levy} and comb generation in silicon nitride microrings with extremely precise comb spacing of 3 $\times$ 10$^{-15}$ with respect to the optical frequency \cite{Foster}, and combs spanning an octave of bandwidth \cite{OkawachiOctave}. Additionally, the flexibility of controlling the FSR of the comb with different resonator dimensions has been established \cite{Johnson}. Most of these investigations have focused on comb generation dynamics (both experimental \cite{HerrUniversal} and theoretical \cite{Matsko,MatskoML}), and on robustness and stability analysis\cite{Del'Haye,Foster} in the frequency domain. Time-domain studies of parametric combs have been reported \cite{Ferdous,Papp,MatskoTransient} and pulses as short as 432-fs have been observed utilizing suitable external phase modulation of individual comb lines. There have been several recent demonstrations that suggest the onset of phase-locking of generated comb lines through the reduction of RF amplitude/phase noise \cite{OkawachiOctave,Johnson,HerrUniversal,MatskoTransient}. However, RF noise reduction or phase locking does not necessarily imply modelocking and ultrashort pulse generation, and time-domain characterization confirming modelocking has not been performed. 

In this paper, we investigate simultaneously the temporal and optical and radio-frequency spectral properties of a parametric frequency combs generated in CMOS-compatible, integrated silicon-nitride microresonators and observe that the system undergoes a transition to a mode-locked state and that ultrashort pulse generation is occurring.  From a 25-nm filtered section of the 300-nm comb spectrum, we observe sub-200-fs pulses at a 99-GHz repetition rate. To illustrate the flexibility of this platform in terms of controlling the pulse repetition rate, we show that by operating with a shorter microresonator, similar duration pulses can be produced at a 225-GHz repetition rate. Calculations indicate that the pulse generation process is consistent with soliton modelocking, which is consistent with very recent work involving comb generation in MgF$_2$ microresonators \cite{KippenbergarXiv}. These results demonstrate that such parametric frequency combs can serve as a source of ultrashort laser pulses that, depending on the pump laser and material system, could produce ultrashort pulses from the visible to the mid-infrared at repetition rates in the GHz to THz regimes.  

\section{Experiment}

\begin{figure}[tp]
\centering\includegraphics[width=10cm]{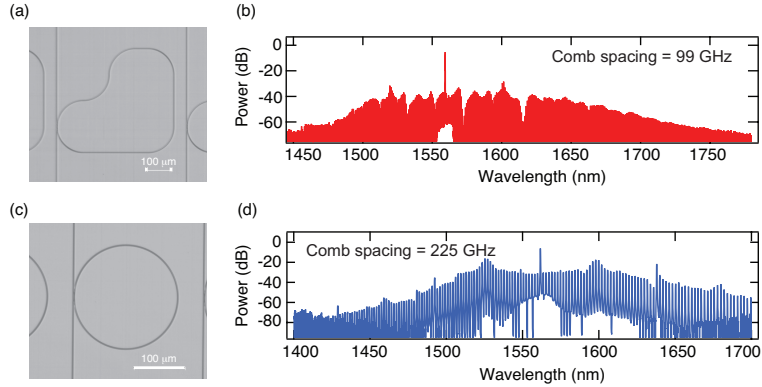}
\caption{(a) A scanning electron micrograph of a silicon-nitride resonator of length 1.44 mm coupled to a bus waveguide. The free-spectral range (FSR) is 99 GHz. (b) Optical spectrum of 99-GHz FSR frequency comb. (c) Micrograph of a silicon-nitride microresonator resonator of radius 112-$\mu$m with an FSR of 225 GHz. (d) Optical spectrum of a 225-GHz FSR frequency comb.}
\label{fig:fig1}
\end{figure}

\begin{figure}[tp]
\centering\includegraphics[width=9cm]{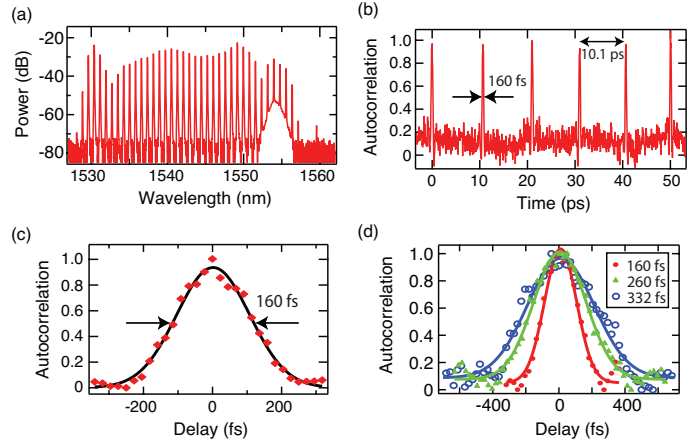}
\caption{(a) Filtered optical spectrum of a 99-GHz FSR frequency comb. The filter bandwidth is 25 nm. (b) Normalized autocorrelation trace of the pulse train obtained from the filtered comb. The pulse separation is 10.1 ps. (c) Zoomed-in view of a single pulse with a 160-fs FWHM pulse width.}
\label{fig:fig2}
\end{figure}

For our experiments, the pump wave is derived by amplifying a single-frequency tunable diode laser at 1560 nm to 2.5 W using an erbium-doped fiber amplifier (EDFA) and coupling it into the bus waveguide using a lensed fiber. Both the microresonator and the bus waveguide are fabricated monolithically in a single silicon-nitride layer, allowing for robust and environmentally stable operation. We use a resonator with a cavity length of 1.44 mm, which corresponds to a 99-GHz FSR. A micrograph image of this resonator is shown in Fig. \ref{fig:fig1}(a). The input polarization is adjusted to quasi-TE using a fiber polarization controller. As the coupled power in the resonator is increased the threshold for parametric oscillation of a signal/idler pair is reached. Further increases in the power coupled into the microresonator lead to cascaded FWM and higher-order FWM processes, resulting in the generation of numerous comb lines. The output is collected using an aspheric lens and sent to a 4-$f$ shaper, which is used as an adjustable wavelength filter. The zero-order signal of the 4-$f$ shaper is sent to an optical spectrum analyzer (OSA) to monitor the entire comb spectrum. The generated frequency comb spanning $>$300 nm is shown in Fig. \ref{fig:fig1}(b). To characterize the pulse dynamics, we filter a 25-nm section (32 comb lines) of the comb centered at 1546 nm and amplify it with an EDFA to sufficiently high power levels to allow for autocorrelation measurements. Our choice of the 25-nm bandwidth corresponds to the bandwidth over which we could amplify the signal for characterization. An OSA trace of the filtered spectrum of the frequency comb is shown in Figure \ref{fig:fig2}(a). If the generated comb lines have a definite phase relationship, then a periodic pulse train in the time domain with a repetition rate given by the comb spacing should be observed. We investigate the temporal properties of the generated comb using an intensity autocorrelator, and Fig. \ref{fig:fig2}(b) shows the normalized autocorrelation trace of the observed pulse train for the filtered comb shown in Fig. \ref{fig:fig2}(a). The amplified output is sent through a length of single-mode fiber (SMF-28) for group-velocity dispersion compensation. The background noise due to amplified spontaneous emission due to optical amplifiers has been subtracted out. The pulse train has a 10.1-ps pulse period, which corresponds to the microresonator FSR of 99 GHz. Figure \ref{fig:fig2}(c) shows a zoomed-in viewgraph of a single pulse. We measure pulse-widths as short as 160-fs (full-width half maximum, FWHM), which is close to the transform limit of 140 fs for the filtered bandwidth, assuming temporal Gaussian pulses. Using the entire comb bandwidth offers potential for generating single-cycle pulses with even shorter pulse-widths (i.e. 12 fs). We estimate the peak power of the pulses to be 1.2 W, and the temporal output is stable as long as the pump wavelength is on-resonance and there is no variation in coupling to the resonator. These results confirm our previous observation that the frequency comb transitions to a low-noise, phase-locked state \cite{OkawachiOctave}, and we address this issue in greater detail below. Passive modelocking has since been observed in MgF$_2$ crystalline resonators \cite{KippenbergarXiv}.

To confirm that we are indeed generating pulses and not incoherent spikes from the microresonator-based frequency combs, we utilize temporal broadening in a dispersive media. We send the amplified 25-nm filtered section of the frequency comb output from the 1.44-mm-long microresonator cavity (corresponding to a 99-GHz FSR) through a few meters of single mode fiber (SMF) and temporally characterize the generated pulses using an intensity autocorrelator as indicated in the main text. Figure \ref{fig:fig2}(d) shows the result of this measurement. We clearly observe temporal broadening of the pulses from 160 fs to 332 fs due to dispersive propagation through the additional lengths of SMF.  

\begin{figure}[bp]
\centering\includegraphics[width=9cm]{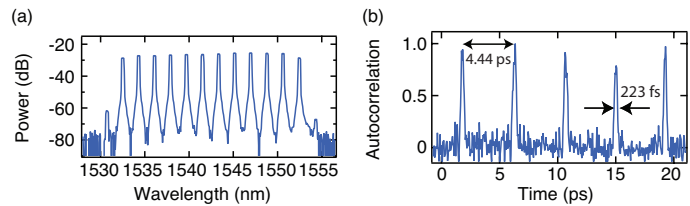}
\caption{(a) Filtered optical spectrum of a frequency comb with 225-GHz free spectral range. (b) Normalized autocorrelation trace of pulse train obtained from the filtered comb. The pulse separation is 4.44 ps and the FWHM pulsewidth is 223 fs.}
\label{fig:fig3}
\end{figure}

The silicon-nitride platform allows for unmatched flexibility in terms of controlling FSR, without changing the cavity dispersion. This allows for design and fabrication of multiple high-repetition-rate pulse sources with a specified pulse repetition rate all on a single chip. As an example, we use a 112-$\mu$m-radius ring-resonator, which corresponds to a 225-GHz FSR [see Fig. \ref{fig:fig1}(c)-(d)]. The optical spectrum of the corresponding comb is shown in Figure \ref{fig:fig1}(d) for the case in which we pump at 1559 nm. After filtering out a 25-nm section of the comb (14 lines) centered at 1543 nm [Fig. \ref{fig:fig3}(a)], we amplify and measure the pulse train with the autocorrelator. We observe uncompressed sub-225-fs pulses at the expected 4.44-ps time interval [Fig. \ref{fig:fig3}(b)], which corresponds to a 225-GHz repetition rate. 

\begin{figure}[tp]
\centering\includegraphics[width=10cm]{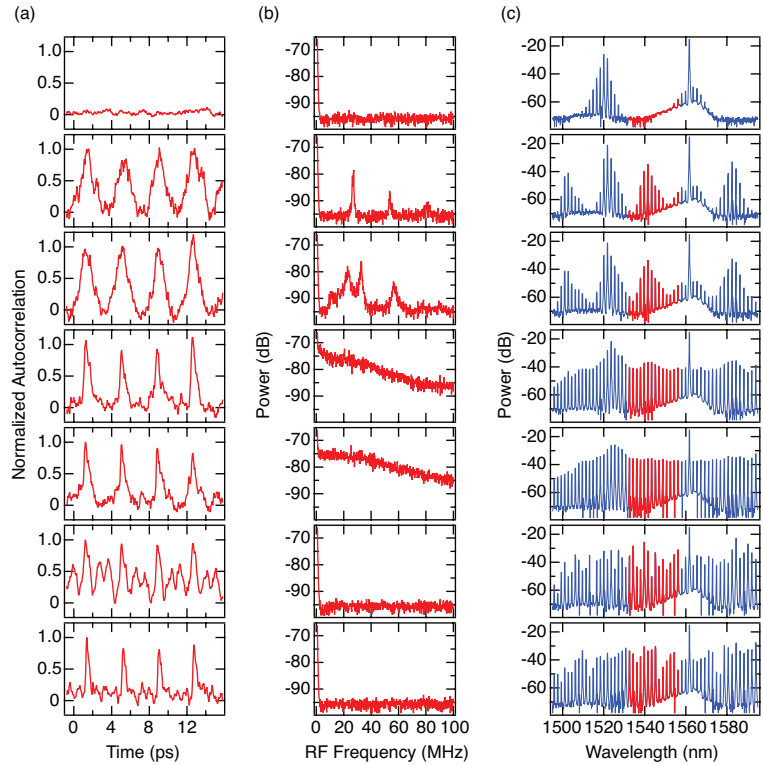}
\caption{(a) From top to bottom, pulse formation dynamics as the laser is tuned into resonance of the microresonator, thereby increasing the power coupled into the microresonator. (b) RF amplitude noise corresponding to each stage of pulse formation shown in column a to its left. (c) Optical spectrum of comb generation dynamics. Full comb is represented in blue, the filtered section of the comb used for pulse generation is shown in red.}
\label{fig:fig4}
\end{figure}

\begin{figure}[tp]
\centering\includegraphics[width=8cm]{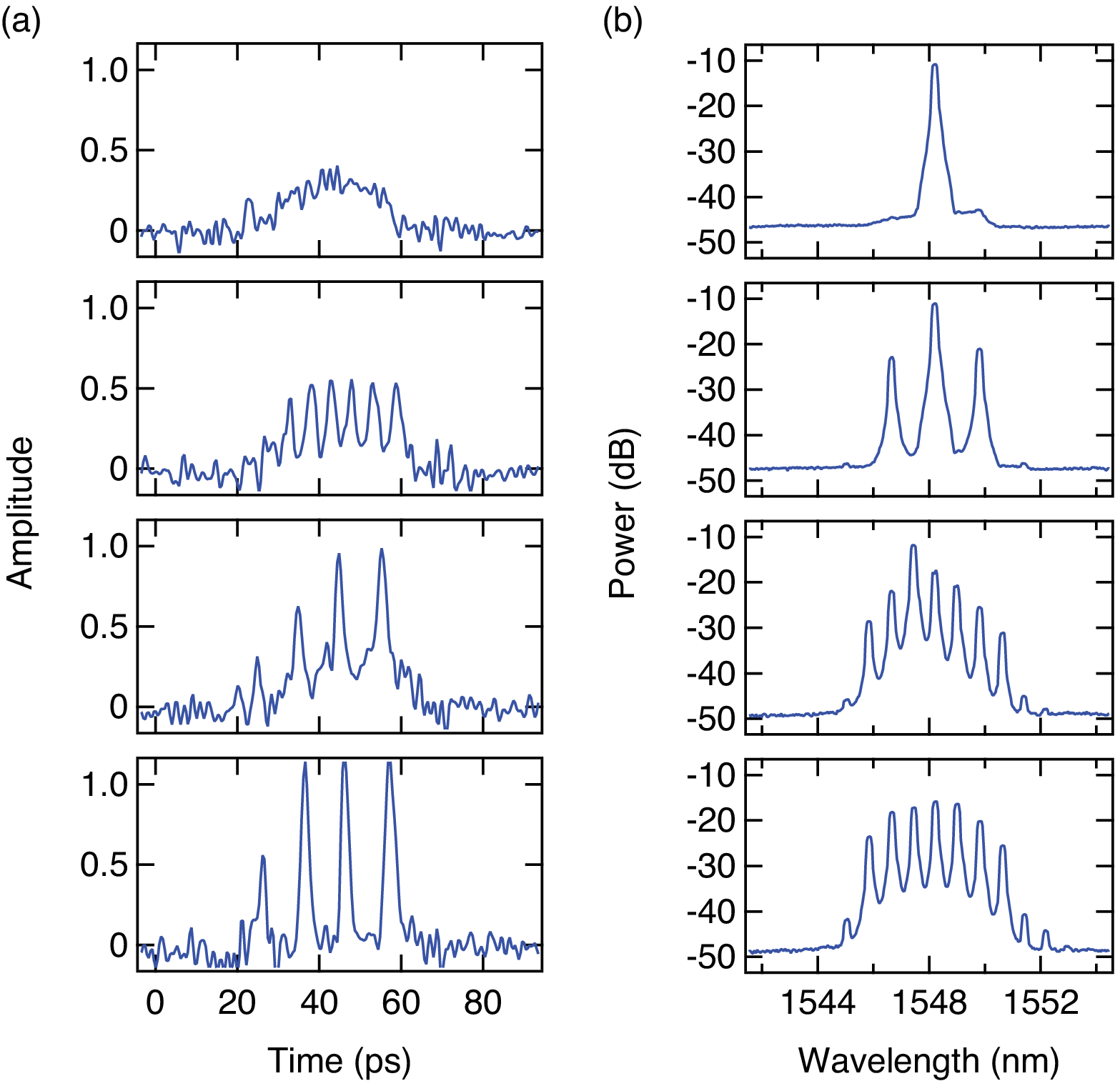}
\caption{(a) Single-shot characterization of temporal evolution (top to bottom) of pulses generated in a microresonator with 99-GHz FSR measured with an ultrafast temporal magnifier and a real-time oscilloscope. (b) Spectral evolution of the generated frequency comb as the pump is tuned into resonance and the power inside the microresonator increases.}
\label{fig:fig5}
\end{figure}

Finally, we investigate the pulse formation dynamics in the parametric comb. Similar to our previous measurements, we filter a 25-nm section of the 225-GHz FSR comb. After splitting the filtered output, we monitor simultaneously the autocorrelation trace, the RF amplitude noise, and the optical spectrum of the generated comb. The results of the measurement are shown in Fig. \ref{fig:fig4}. The leftmost column shows the autocorrelation traces of generated temporal waveforms [Fig. \ref{fig:fig4}(a)], the middle column [Fig. \ref{fig:fig4}(b)] shows the RF amplitude noise spectrum measured with an RF spectrum analyzer, and the rightmost column shows the optical spectrum [Fig. \ref{fig:fig4}(c)] of the generated frequency comb corresponding to each stage of pulse formation as the pump wavelength is tuned into the microcavity resonance (top to bottom). The 25-nm filtered section of the frequency comb that is used for temporal characterization is represented in red color in each optical spectrum trace. As we tune the pump wavelength into resonance and more power is coupled into the microresonator, cascaded FWM takes place and comb lines are generated several FSRs away from the pump where the cavity modes experience the maximum FWM gain due to the interplay between the group velocity dispersion and nonlinearity. Similar comb generation dynamics in other platforms have also been observed \cite{HerrUniversal}. Gradually, small clusters of comb lines (mini-combs) begin to appear centered at each of the cascaded FWM peaks. Simultaneously, a sinusoidal pulse train appears, and several low-frequency peaks are observed in the RF domain. Further tuning into resonance leads to gradual equalization of the amplitude of the comb lines and reduction of pulse duration. However, at the same time, the number of peaks in the RF domain increases, and the linewidth of each peak broadens until it becomes a broad plateau. We attribute these peaks to beating between adjacent mini-combs due to the fact that, at this point in the generation process, the mini-combs are uncorrelated and could have different comb spacings and/or dc offsets \cite{HerrUniversal}. As the comb lines equalize, the spectral overlap between the mini-combs becomes more extensive, resulting in a significant increase in the RF noise. However as we tune deeper into the resonance, the comb reaches a sudden transition stage where the RF noise suddenly decreases by $>$20 dB, which is a signature of phase-locking of the comb akin to modelocking in a femtosecond laser. The temporal waveform at this point is very sensitive to the amplitude of the individual comb lines, and we observe modulations in both the autocorrelation and the optical spectrum. Finally, as the pump is tuned further into resonance, we again observe equalization of the comb line amplitudes, and pulses with the shortest durations are generated. 

We further confirm that the temporal waveforms obtained from the intensity autocorrelation using the 99-GHz FSR comb are indeed coherent, mode-locked pulses and not coherence spikes utilizing a temporal magnifier. This is an independent measurement technique based on time-lens technology that allows for single-shot characterization of ultrafast temporal waveforms using a GHz-bandwidth real-time oscilloscope through temporal stretching of the input waveform \cite{Salem,OkawachiTL}. We send a 7-nm filtered section of a 99-GHz-FSR frequency comb to the time-magnifier. The temporal magnifier system outputs temporally stretched snap-shots of the input waveform, which are recorded using a real-time oscilloscope. Figures \ref{fig:fig5}(a) and \ref{fig:fig5}(b) (from top to bottom), respectively, show the temporal and spectral evolution of a frequency comb generated using a 99-GHz-FSR microresonator. We observe that, as the pump wavelength is tuned into resonance and a single comb line is generated, there is a cw temporal output. When the pump is tuned further into resonance, we observe three comb lines in the spectral domain, corresponding to a pulse repetition rate that is twice the FSR of the microresonator. As the pump is further tuned in, we see that every adjacent comb line is filled in. However, at this point, in the time domain, we observe a distorted pulse train with a significant background, indicating that the pulse train is not phase-locked. Finally, as we tune into the deepest point of the resonance, we observe narrow pulse-width with no background, indicating the onset of modelocking. These results clearly validate the existence of coherent and mode-locked pulses generated from microresonators pumped with a single-frequency cw pump. 

To support the claim that this system exhibits soliton modelocking behavior similar to that in laser systems, we calculate the nonlinear length $L_{NL} = 1/\gamma P$ and dispersion length $L_D = \tau_p ^2 /| \beta _2|$ within the resonator for the pulses generated from a 100-nm bandwidth of the 99-GHz-FSR comb, where $\lambda=1560$ nm is the center wavelength, $\gamma=$1.09 W$^{-1}$m$^{-1}$ is the nonlinear parameter, $\tau_p=14.5$ fs is the pulse duration, and $\beta_2=-0.064$ ps$^2/$m is the group-velocity dispersion.  The peak power $P=633$ W for determining $L_{NL}$ of the 100-nm section for this calculation is estimated based on the peak power of the filtered 25-nm section of the comb and by taking into account the spectral shape of the comb excluding the pump. The estimated average power circulating in the ring is consistent with the theoretically calculated value including all possible losses. The calculated soliton number $N=\sqrt{L_D/L_{NL}}$ is 1.5, which strongly suggests that the system is operating under soliton modelocking conditions. Although there have been previous investigations for modelocking in microresonator-based combs \cite{MatskoML,MatskoHardSoft}, the underlying mechanism in our system remains unknown. One possible mechanism could be that the relatively high optical intensities circulating inside the cavity lead to local changes in the nonlinear refractive index resulting in greater spatial confinement of the pulse and reduced losses since there is less absorption at the core-cladding interface. Alternatively, the pulse train may be produced due to the formation of temporal cavity solitons \cite{Leo}, where contributions from dispersion and loss are compensated by nonlinearity and a coherent driving beam \cite{Leo}, and recent experiments using MgF$_2$ microresonators support this conclusion \cite{KippenbergarXiv}.

\section{Conclusion}
We demonstrate the first observation of sub-200-fs pulses from an on-chip, silicon-based frequency comb source. We observe that as the frequency comb develops, a transition occurs into a stable, low-noise phase-locked state similar to that which occurs in conventional mode-locked femtosecond laser sources. This demonstration represents a significant advancement towards the development of integrated, stabilized, and compact ultra-high repetition-rate femtosecond sources. Furthermore, since the FSR and the cavity dispersion can be independently controlled, this system allows for building numerous ultrafast sources on the same chip with unparalleled flexibility in repetition rates and operating wavelengths from the near visible to the mid-infrared.

We acknowledge support from Defense Advanced Research Projects Agency (DARPA) via the QuASAR program and the Air-Force Office of Scientific Research. This work was performed in part at the Cornell Nano-Scale Facility, a member of the National Nanotechnology Infrastructure Network, which is supported by the National Science Foundation (NSF) (grant ECS-0335765). The authors also acknowledge useful discussions with Jaime Cardenas.
\end{document}